\documentclass[pdflatex,sn-mathphys-num]{sn-jnl}

\usepackage{graphicx}%
\usepackage{multirow}%
\usepackage{amsmath,amssymb,amsfonts}%
\usepackage{amsthm}%
\usepackage{mathrsfs}%
\usepackage[title]{appendix}%
\usepackage{xcolor}%
\usepackage{textcomp}%
\usepackage{manyfoot}%
\usepackage{booktabs}%
\usepackage{algorithm}%
\usepackage{algorithmicx}%
\usepackage{algpseudocode}%
\usepackage{listings}

\begin{document}

\title[Article Title]{Dynamic control of laser driven electron acceleration in a photonic structure using programmable optical pulses}


\author*[1]{\fnm{Sophie} \sur{Crisp}}\email{sophiecrisp@physics.ucla.edu}

\author[2]{\fnm{R. Joel} \sur{England}}

\author[1]{\fnm{Alexander} \sur{Ody}}

\author[1]{\fnm{Pietro} \sur{Musumeci}}\email{musumeci@physics.ucla.edu}

\affil*[1]{\orgdiv{Physics and Astronomy}, \orgname{University of California, Los Angeles}, \orgaddress{\street{Portola Plaza}, \city{Los Angeles}, \postcode{90095}, \state{California}, \country{USA}}}

\affil[2]{\orgname{SLAC National Accelerator Laboratory}, \orgaddress{\street{2575 Sand Hill Rd}, \city{Menlo Park}, \postcode{94025}, \state{California}, \country{USA}}}

\abstract{
Progress in optical techniques has made precision control of the phase profile in optical pulses common and accessible in scientific laboratories. Carefully shaping the field profile of a laser pulse can be used to master the dynamics of electrons traveling in photonic accelerating structures such as the ones obtained by precisely aligning two dielectric gratings. Here we show that by applying a liquid-crystal-mask to program the phase and amplitude of an infrared laser pulse in combination with a pulse front tilt scheme, it is possible to implement dynamic control on a laser accelerator. This results in a nearly limitless live tuning capability of the accelerator beam dynamics, allowing the demonstration software-based correction of structure and optical front imperfections, implementation of transverse focusing schemes, control of the energy and charge of the output beam, and ultimately optimization of the interaction length, leading to measured energy gains of up to 0.55~MeV.}






\maketitle

\section{Main}\label{sec:main}

Particle accelerators have become a mainstay of modern science and technology. High energy colliders enable the discovery of new particles, while synchrotrons and free-electron lasers are crucial tools for the next generation of chemistry, biology and material sciences \cite{shiltsev2021modern, pellegrini2016physics}. In addition, a large number of medical and industrial applications of relativistic beams continue to drive the development of compact and cost-effective solutions \cite{hanna2012rf, cleland2006industrial}. Conventional accelerators use radiofrequency fields on the order of 10-100~MV/m to accelerate particles, but the cavities which support these fields are large and expensive, as are the power sources. With the advent of dielectric materials that can withstand GV/m fields, higher gradient photonic structures that can be mass-produced using modern nanofabrication techniques have become of great interest \cite{England:RMP}. Due to the extremely small size of the accelerating channels, GV/m field intensities can be achieved using modest ($\mu$J to mJ) pulse energies from widely available compact laser sources. Not only do these dielectric laser accelerators (DLAs) promise to shrink accelerator dimensions, fabrication and operation costs by several orders of magnitude, they also enable the generation of electron bunches with sub-nm normalized emittances and attosecond characteristic time scales \cite{Schonenberger2019, Black2019_netaccel}.

A key challenge in using lasers to directly drive dielectric accelerator structures lies in the ability to manipulate the electromagnetic field to control the transverse and longitudinal dynamics of charged particles \cite{shiloh2022}. Precise tuning of the phase of the field experienced by the electrons is essential for maximizing particle trapping, ensuring prolonged resonant interaction, and controlling the transverse dynamics. In experimental efforts, the initial demonstration of energy modulation mediated by simple gratings quickly gave way to the demonstration of higher gradients, longer interactions, and the introduction of on chip power delivery and focusing elements \cite{Peralta2013,Breuer2013, Cesar2018_highintensity, Cesar2018PFT,sapra2020_science}. Different materials and drive laser wavelengths have also been explored \cite{zheng2023efficiently, bar2014plasmonic, mei2023dielectric, bruckner2024mid}. In order to mitigate the large defocusing fields which accompany high acceleration gradients, alternating phase focusing (APF) and ponderomotive focusing schemes have been proposed \cite{Niedermayer2018_APFconcept, Naranjo2012_Ponderomotive}, and demonstrated \cite{Shiloh2021_firstAPFdemo,Zhang2024_APF_cherenkovdla}. In all previous implementations, phase control has been experimentally achieved either by manipulating multiple independent laser pulses or breaking the translational symmetry of pillar structures by varying the period or adding physical phase-resetting gaps \cite{Niedermayer2021_attosecond, Chlouba2023_coherent,Broaddus2024_subrelAPF}. 

\begin{figure}
    \centering
    \includegraphics[width = \textwidth]{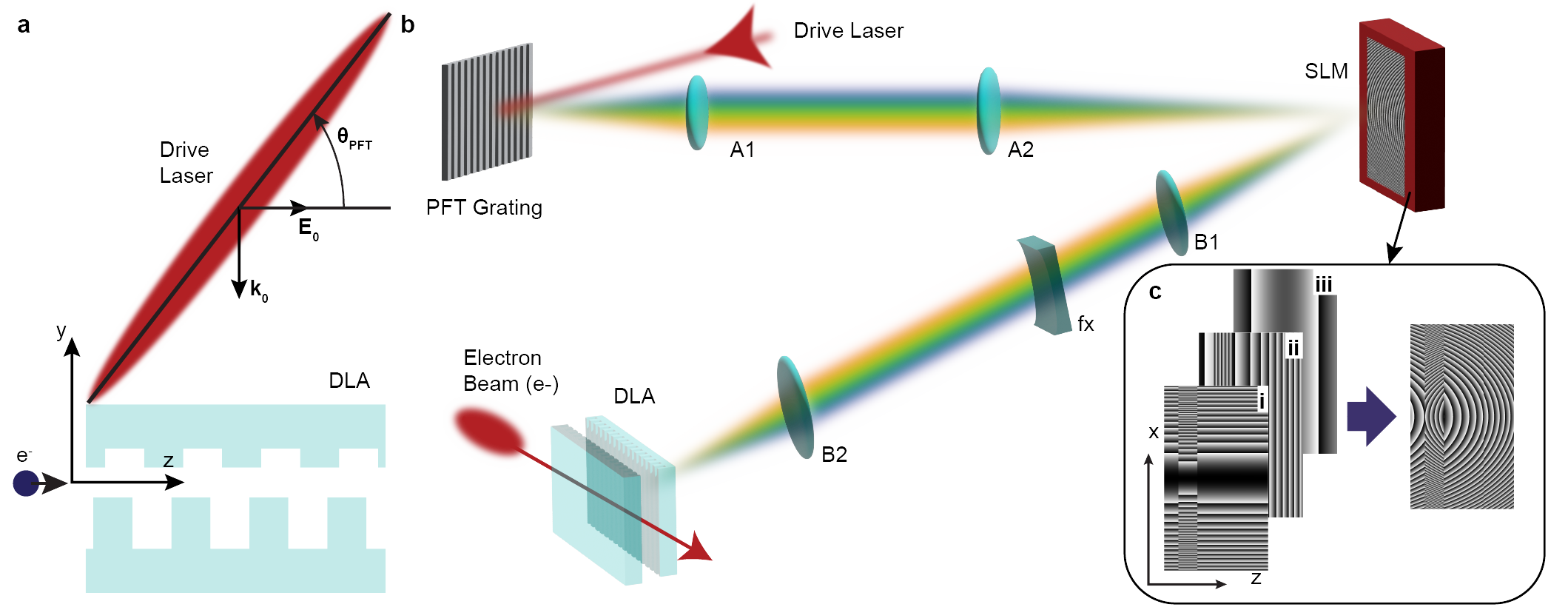}
    \label{fig:Concept}
    \caption{Conceptual design of a DLA accelerator with programmable phase control \textbf{a} Cartoon drawing of the pulse-front-tilted laser incident on an asymmetric fused silica dual grating structure. \textbf{b} Optical system used to generate arbitrary phase profiles on a laser with PFT. Imaging planes are located at the PFT grating, SLM, and DLA, such that PFT is transferred to the DLA and phase profiles in the electron trajectory direction ($z$) are transferred temporally to the electron beam. A cylindrical lens between the SLM and DLA allows for decoupling between $x$ (transverse to the electrons) and $z$ so that the SLM can be used simultaneously for phase and amplitude control. \textbf{c} Example phase masks that can be programmed onto the SLM to manipulate the electron beam. The final applied mask (right image) is the sum of (i) a z-dependent fresnel lens, (ii) a phase matching mask, and (iii) a nonlinear effect compensation mask.}
\end{figure}

Meanwhile, spatial light modulators (SLM) have become critical tools in quantum optics, microscopy and holography, \cite{Yiqian2023_SLMReview, Yao2006_SLMentanglement, Maurer2011_SLMMicroscopy, Grier2003_SLMTweezers} due to their off-the-shelf capability to provide programmable nearly arbitrary shaping of the phase and/or amplitude of laser pulses. In this paper, we leverage the versatility offered by liquid-crystal based SLM to demonstrate tunable control of the acceleration dynamics in a DLA based on a simple fused silica dual-grating structure illuminated with a pulse-front-tilted (PFT) laser beam. (Fig. \ref{fig:Concept}a)\cite{cesar_2018_allopticalcontrolproposal}. Injecting the drive laser at $\approx 90$ degrees with respect to the propagation axis of the electrons allows us to lengthen the temporal overlap region of laser and electrons and at the same time to project the tunable spatial modulation onto the electrons as they travel through the structure, as illustrated in Fig. \ref{fig:Concept}a. This approach enabled extending the interaction to multi-mm lengths, testing of transverse and longitudinal phase space dynamics control schemes, and led to the demonstration of up to 0.55~MeV energy gains.


Limiting initially the discussion to longitudinal dynamics, the energy modulation in a DLA can be written as a simple integral of the field experienced by the electrons along their trajectory, $\Delta{E} = \mathfrak{Re} [q\kappa \int^{L/2}_{-L/2} E_0(z, t) e^{i\phi(z)} dz]$. The structure factor, $\kappa$, measures how efficiently the structure converts the incident field into accelerating gradient; $L$ is the structure length, and $E_0(z, t)$ represents the laser field amplitude. Maximum energy modulation corresponds to the case when the phase $\phi(z)$ is constant along the interaction such that the resonant phase matching requirement is met, i.e. the electrons do not slip in the accelerating wave when traveling in the structure. In order to appreciate the level of control on the accelerator enabled by the SLM, we can identify the different contributions to the total laser phase $\phi(z) = \phi_{TW}(z) + \phi_{SLM}(z) + \phi_{NL}(I_0, z)$ as the traveling wave, SLM-controllable term and non-linear contribution respectively. The traveling wave phase is defined as $\phi_{TW} = k_g z - \frac{k_0 z}{\beta(z)}$ where $\beta(z)$ is the normalized electron velocity along the structure and $k_g$ and $k_0$ are the wavenumbers of the grating and the laser respectively. In all simulations and experiments described in this paper, these wavenumbers are equal and constant -- the gratings employed to form the structure are simple untapered transmission gratings with a periodicity equal to the drive laser wavelength (i.e. $k_0 = k_g$).  For our relativistic beams the initial normalized velocity value $\beta_0$ is near 1, and phase matching can be obtained varying the incident angle, $\theta_I$, of the laser on the structure or equivalently adding a linear phase ramp $\phi_{SLM} = k_0 \sin(\theta_I) z$ with the SLM; this additional phase term tunes the phase velocity of the accelerating wave and can be used to match it to the electron velocity in the structure \cite{Crisp2021}. As a result of the the DLA-induced energy gain, the velocity of the particles $\beta(z)$ can change along the interaction, which can be compensated by a $z$-dependent slope $\theta_I(z) = \theta_{I0} + \Delta \theta(z)$, where in the small angle approximation $\theta_{I0} \simeq \frac{1}{\beta_0} - 1$. As an example, for 8~MeV ballistic (i.e. constant velocity) electrons the initial matching angle $\theta_{I0} = 1.73$ mrad while $\Delta\theta(z)$ can be used as a tuning parameter to compensate for DLA acceleration-induced dephasing with the SLM phase. Additionally, the use of slab transmission gratings in conjuction with high-intensity laser pulses results in Kerr effect dephasing $\phi_{NL} = n_2I_0(z)k_g\Delta y$. Here, $\Delta y$ refers to the thickness of the bulk fused silica (1~mm), $n_2$ is the nonlinear index of refraction, and $I_0(z)$ is the intensity profile of the laser. Both dephasing contributions due to the acceleration and due to Kerr effects have been observed to strongly limit the interaction length in previous DLA experiments \cite{crisp2024extended, Cesar2018_highintensity}. While other approaches hard-code into the structure the phase variation needed to keep the acceleration in resonance, this is accompanied by significant flexibility limitations \cite{Chlouba2023_coherent,Broaddus2024_subrelAPF,Shiloh2021_firstAPFdemo,niedermayer2020three}. 

In our scheme, we utilize a single laser drive illuminating a DLA structure orthogonally with respect to the electron beam propagation direction (Fig. \ref{fig:Concept}a). A pulse-front-tilt is applied to the laser pulse to extend the interaction \cite{wei2017dual}, which would otherwise be limited to the 100~fs laser pulse length. Programmable control of the optical field seen by the electrons is accomplished by using a reflective 2D liquid crystal array between the grating and the DLA (see methods). In the $z$-dimension (parallel to the electron propagation) the phase applied by the SLM is relay-imaged to the DLA.  The SLM tunability can be used to directly apply on the incoming laser arbitrary phase profiles $\phi_{SLM}$ in order to flatten the phase experienced by the electrons and satisfy the resonant condition. Sample phase masks corresponding to the different contributions to compensate the acceleration via $\theta_I(z)$ and compensate non linear dephasing $\phi_{NL}$ are shown in Fig. \ref{fig:Concept}c,ii/iii. Note that due to the $2\pi$ limit on phase retardation imposed by the SLM, the applied masks are not simple smooth profiles. For instance, a linear tilt across the SLM is generated according to $\text{mod}(k_0 z \sin(\theta), 2\pi)$, so that larger angles result in an increasing number of phase jumps across the SLM screen. 

In addition to phase control, the scheme allows to use the same single programmable LCM element to manipulate the amplitude of the field seen by the electrons. This is accomplished inserting a cylindrical lens (labeled $fx$ in Fig. \ref{fig:Concept}b) to break the imaging condition in $x$ such that applying a variable Fresnel lens profile $\phi_F(x,z) = -\frac{k_0}{2}\frac{x^2}{f_x(z)}$ at the SLM plane results in direct control over the laser transverses size and hence the field intensity at the DLA channel. An example of such mask profile is shown in Fig. \ref{fig:Concept}c,i. Crucially, the accelerator tuning enabled by the SLM extends beyond simple resonant phase matching. As detailed later, the full control of the 2D liquid crystal array allows for advanced phase and amplitude shaping, which can be used to manipulate transverse beam dynamics and engineer regions of strong energy modulation followed by low-field drift sections along the structure. This can be exploited both to locally optimize the acceleration in a particular region of the structure as well as enhance the beam microbunching process to improve the capture and output beam quality.  

\section{Results}\label{sec:results}

We experimentally demonstrate programmable control of DLA acceleration using the 8.7~MeV electron beam from the UCLA Pegasus photoinjector (see Sec. \ref{sec:methods_electrons}) \cite{Alesini2015}. The structure is an asymmetric dual grating structure formed by precisely aligning a pair of custom fabricated transmission gratings to a 1.2 $\mu$m gap (see Sec. \ref{sec:methods_structure}) where the structure factor is $\kappa = 0.05$ \cite{crisp_asymmetricDesign}. The primary observable is the energy spectrum of the electron beam after it has passed through the structure and interacted with the laser field. Laser on and laser off electron spectra are then recorded on either the wide acceptance or high resolution spectrometer screens (see Methods and refer to Fig. \ref{fig:supplementarySchematics}) and compared to extract the maximum energy gain or loss experienced by the particles.

\begin{figure}
    \centering
    \includegraphics[width = \textwidth]{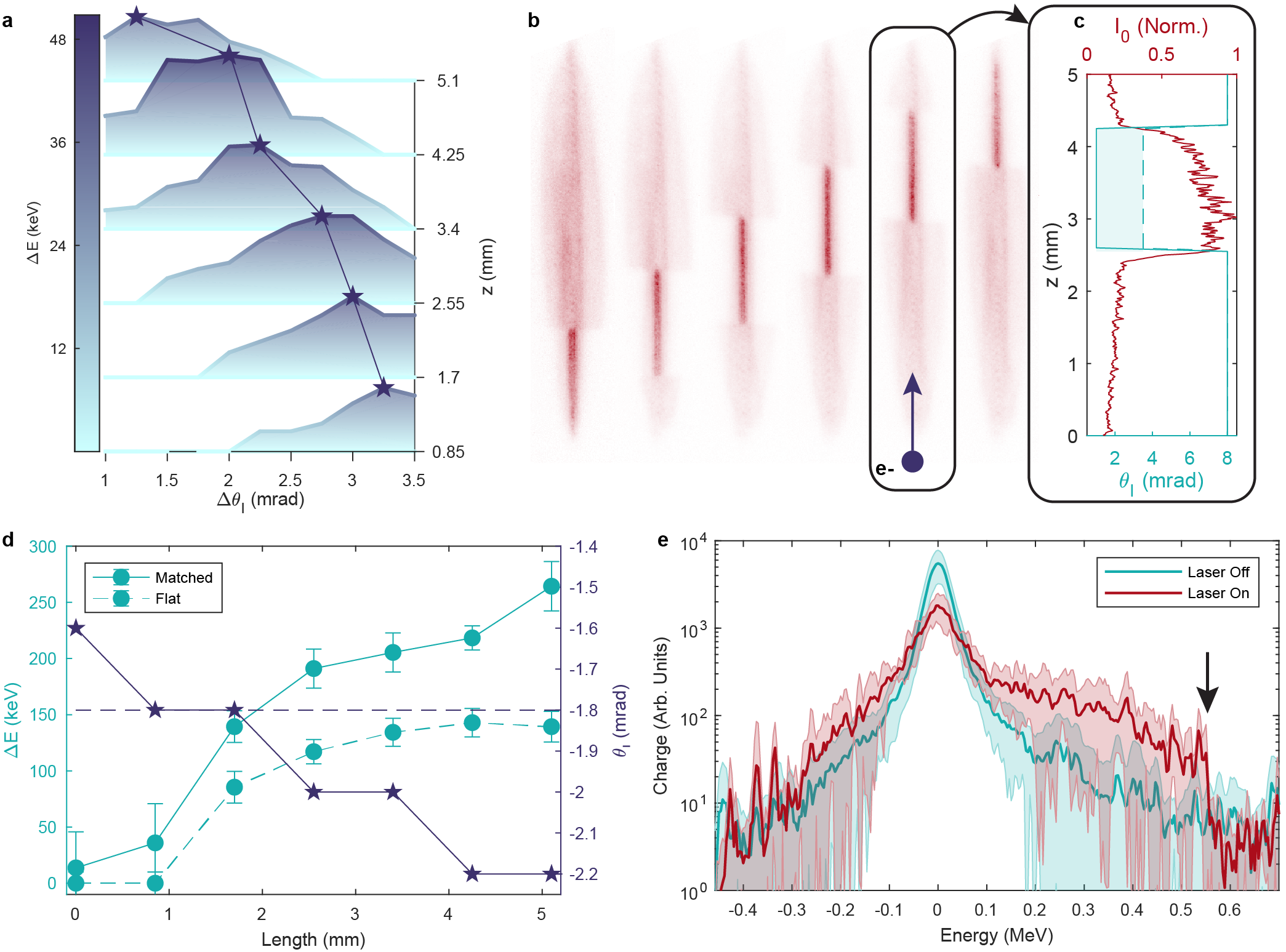}
    \caption{Phase Tuning \textbf{a} Maximum energy gain at different linear phases along the structure. Each lineout corresponds to the illumination of 1.7~mm of structure, centered about the designated z location. The rest of the laser is defocused and $>$3~mrad detuned from resonance. \textbf{b} Overhead images of laser profiles as seen on phosphor screen. Each laser profile corresponds to a lineout of (\textbf{a}). \textbf{c} Corresponding phase mask and amplitude profile to the illumination of the central 1.7~mm of the structure. \textbf{d} Maximum energy gain as a function of illumination length, comparing the curved, tuned phase profile to a linear tilt. \textbf{e} Electron spectrum for the largest obtained energy modulation.}
    \label{fig:phase_basic}
\end{figure}

In order to best phase match to the accelerating electrons, the SLM phase profile is first optimized in segments, varying the linear tilt (or phase ramp) in 1.7~mm sections of the DLA and recording the measured change in the electron beam energy $\Delta E$ as a function of the applied $\Delta\theta_I$ as shown in Fig. \ref{fig:phase_basic}\textbf{a}). In this way, we can exceed the interaction length limitations of the simpler flat-phase profile as used by Crisp et al. \cite{crisp2024extended}. In areas designated for minimal interaction, we deliberately introduce a $\Delta\theta=3$~mrad offset, ensuring that the field in that portion of the DLA remains non-resonant with the electrons. In addition, the intensity is reduced by more than a factor of 4 in these same regions by applying a variable focal length Fresnel lens in the $x$-direction. Corresponding laser profiles on a phosphor screen adjacent to the DLA are viewed with an overhead camera and reported in Fig. \ref{fig:phase_basic}\textbf{b}. In the right panel, we show the extracted amplitude and applied phase masks on such laser, indicating that any observed energy gain can be attributed only to the high intensity region. Each star in Fig. \ref{fig:phase_basic}a corresponds to the optimal $\Delta\theta_I$ at that structure location. We then construct a complete phase map $\theta_I(z)$ by stitching together the locally optimal $\Delta\theta_I$ values along the structure.

The benefits of this approach are clearly shown by comparing the energy gain as a function of the DLA accelerator length in Fig.\ref{fig:phase_basic}d (note that the optimal $\Delta\theta_I$ vs. z curve was observed to be sensitive to day-to-day small changes in e-beam alignment, energy and laser setup so the data in this plot corresponds to a slightly different tune than a). The interaction length is adjusted as before by varying the Fresnel lens, but in this case we illuminate increasing lengths of the structure with high intensity with each step. The SLM is set to apply the optimized matched phase profile. The result is a marked increase in the energy gain compared with the flat-phase case, corresponding to a longer (up to 4~mm) phase-matched interaction over the grating structure. 


Noticing an approximately linear variation of the optimal incident angle over the structure length, we found that by properly adjusting the lens positions in the optical system, we could obtain an incoming quadratically curved phase front that would remove the $z$-dependence in the SLM-applied $\theta_I(z)$ map. In this configuration, it became possible to replace the SLM with a piezo-controlled mirror to maintain the capability to fine-tune to the resonant incident angle $\theta_I0$. Overcoming the severe transmission losses and fluence limitations associated with the use of the LCM enabled us to maximize the intensity incident on the DLA accelerator. With the piezo mirror in place and the lenses tuned to flatten the $\theta_I(z)$ map, at an incident field of 6~GV/m, a DLA-record peak energy gain of $0.55\pm0.05$~MeV (see Fig. \ref{fig:phase_basic}e) was observed. 

\subsection{Advanced phase and amplitude manipulations for transverse and longitudinal dynamics control}
\label{sec:Transverse}
With the implementation of the SLM scheme, accelerator tuning is not limited simply to resonant condition matching and we can explore the application of LCM-based phase and amplitude control to the study of different transverse focusing schemes. Due to the micron-scale aperture, the transverse acceptance of a DLA is inherently limited, which is often stated as one of the main drawbacks of this advanced accelerator technique. To make matters worse, strong defocusing forces ($\propto\kappa k_0 E_0$ or MT/m) are present when the particles are accelerated at phases where the longitudinal motion is stable \cite{England:RMP}. Nevertheless, it has been pointed out recently that proper shaping of the DLA fields themselves can lead to particle confinement\cite{Niedermayer2018_APFconcept, Naranjo2012_Ponderomotive}. Multiple approaches have been proposed, generally taking advantage of the fact that at phases different than the accelerating phase the DLA fields are strongly focusing. So far, the realization of these schemes required building custom nanofabricated structures with carefully designed phase jumps. This restricts the range of usable beam energies and input laser intensities, and is not fully scalable to larger structures as manufacturing and laser profile errors stack up \cite{Shiloh2021_firstAPFdemo}. Leveraging the live control offered by the SLM, we employ our setup to test and compare two of such approaches, namely ponderomotive focusing and alternate phase focusing (APF). In the experiment, due to the pointing jitter and mismatched transverse phase space of the incoming e-beam, particle throughput is highly variable, confounding any measurement of charge transmission. Nevertheless, there are clear signatures of these schemes in the energy spectra of the beams, providing insights into the underlying beam dynamics. 

\begin{figure}
    \centering
    \includegraphics[width = \textwidth]{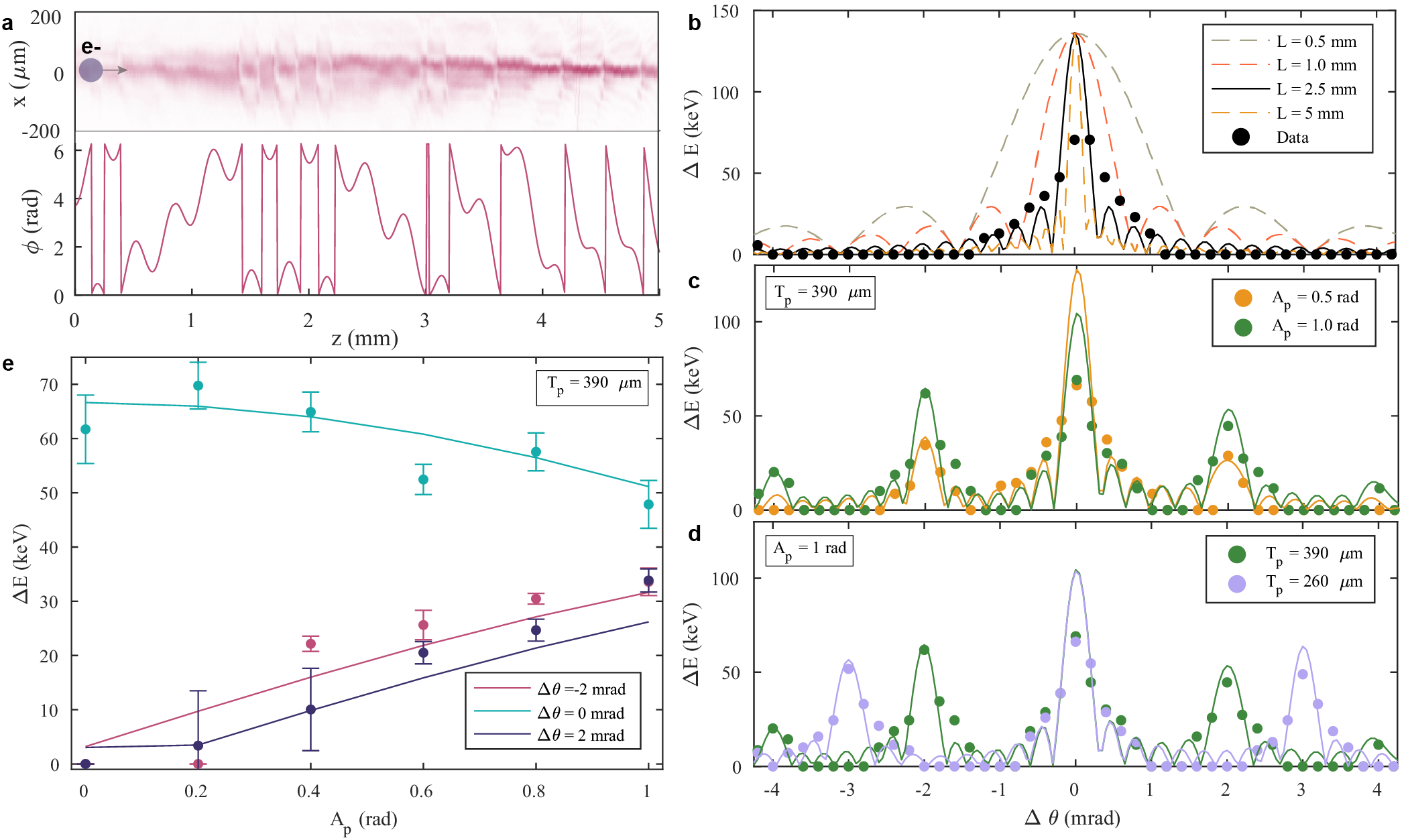}
    \caption{Ponderomotive Excitation of a DLA accelerator \textbf{a} Laser profile (upper) and corresponding phase profile (lower) across the structure for the $\Delta\theta = -2$~mrad, $A_p = 1.0$~rad, $T_p = 390~\mu$m case. \textbf{b} The laser is tuned to the matched laser phase, and then the linear phase tilt $\Delta\theta$ is scanned using the SLM over a $\pm$ 5 mrad range. Several simulation traces are plotted corresponding to the prediction of a simple ballistic model for the DLA acceleration over different interaction lengths $L$. \textbf{c},\textbf{d} Energy modulation as a function of $\Delta\theta$, for a variety of ponderomotive phase contributions. In \textbf{c}, $T_p = 390 \mu$m. In \textbf{d}, $A_p = 1$~rad.\textbf{e} Maximum energy modulation for the three largest spatial harmonics (at $\Delta \theta$ = 0 and $\pm$ 2 mrad) respectively. }
    \label{fig:Ponderomotive4}
\end{figure}

Ponderomotive focusing relies on exciting different spatial harmonics in the structure so while the resonant harmonic is used for acceleration, the non resonant spatial harmonic provides an average transversely focusing force \cite{Naranjo2012_Ponderomotive}. In practice, multiple spatial harmonics can be realized by adding to the phase mask a sinusoidal modulation with amplitude $A_p$ and period $T_p$. By the Jacobi-Anger expansion $e^{iA_p \sin(\frac{2\pi{z}}{T_p})} = \sum J_n(A_p) e^{i\frac{2\pi n}{T_p}z}$, this phase modulation results in a spectrum of spatial harmonics with different phase velocities. The phase lineout along the DLA (lower panel) and the corresponding laser intensity distribution (upper panel) when this oscillatory phase contribution is superimposed to the matched quadratic phase mask are shown in Fig. \ref{fig:Ponderomotive4}a. The measured intensity profile of the laser highlights the need to minimize the SLM phase discontinuities since they translate into undesirable gaps in intensity (see Methods \ref{sec:methods_SLM})\cite{moser2019model}.  

By adding a constant $\Delta \theta$ offset to the applied phase mask, it is possible to measure the energy modulation as a function of the phase velocity detuning and visualize the spatial harmonics in the accelerating wave as they become resonant with the beam. When no phase modulation is added, there is only one peak at $\Delta \theta$ = 0 with a width inversely proportional to the interaction length (Fig. \ref{fig:Ponderomotive4}\textbf{b}). A comparison of the peak width with a ballistic model of the interaction yields an estimate for the effective interaction length as $2.5\pm0.5$~mm. In Fig. \ref{fig:Ponderomotive4}\textbf{c} we report the result of the same measurement when ponderomotive sinusoidal phase modulations with different $T_p$ and $A_p$ are applied. Up to the n = $\pm 2$ modes are clearly visible, especially for larger modulation amplitudes. For $|n|>3$ the resonant energy gains approach the noise floor; this is also partly due to the diffraction efficiency of the SLM greatly decreasing at larger phase slopes. We see that the positions of the excited modes at $\Delta\theta = n k_p / k_g$ agree perfectly with the expectation (\ref{fig:Ponderomotive4}d). In Fig. \ref{fig:Ponderomotive4}e we also show that the relative amplitude of the peaks at the $n = 0,\pm 1$ harmonics matches the predicted behavior $J_n(A_p)$ from the Jacobi-Anger expansion. A larger phase modulation amplitude results in more power and therefore larger energy exchanges for the $n \neq 0$ modes. In practice, the choice of $A_p$ must be made on the basis of necessary focusing power for stable acceleration \cite{Naranjo2012_Ponderomotive}. 

Simulations indicate that ponderomotive focusing can be very effective at confining the particles to the accelerating gap, and our data demonstrates that this approach is well suited to be implemented with the SLM, but the transverse confinement comes at the expense of energy gain; for example, when using the $n = -1$ mode for acceleration, the vast majority of the input energy will go to the $n = 0$ mode for focusing and the electron beam is therefore accelerated only at a relatively small fraction of the maximum gradient.
As proposed in previous designs ~\cite{cesar_2018_allopticalcontrolproposal, Naranjo2012_Ponderomotive} and also in the supplementary material of this paper, stable acceleration conditions require ponderomotive phase oscillation amplitudes $A_p$ between 0.2 and 0.6, thus achieving utilizing less than 10-30 \% of the field amplitude for acceleration. If one wants a higher effective acceleration gradient while still confining the beam, an alternate phase focusing (APF) scheme can be used.

\begin{figure}
    \centering
    \includegraphics[width = \textwidth]{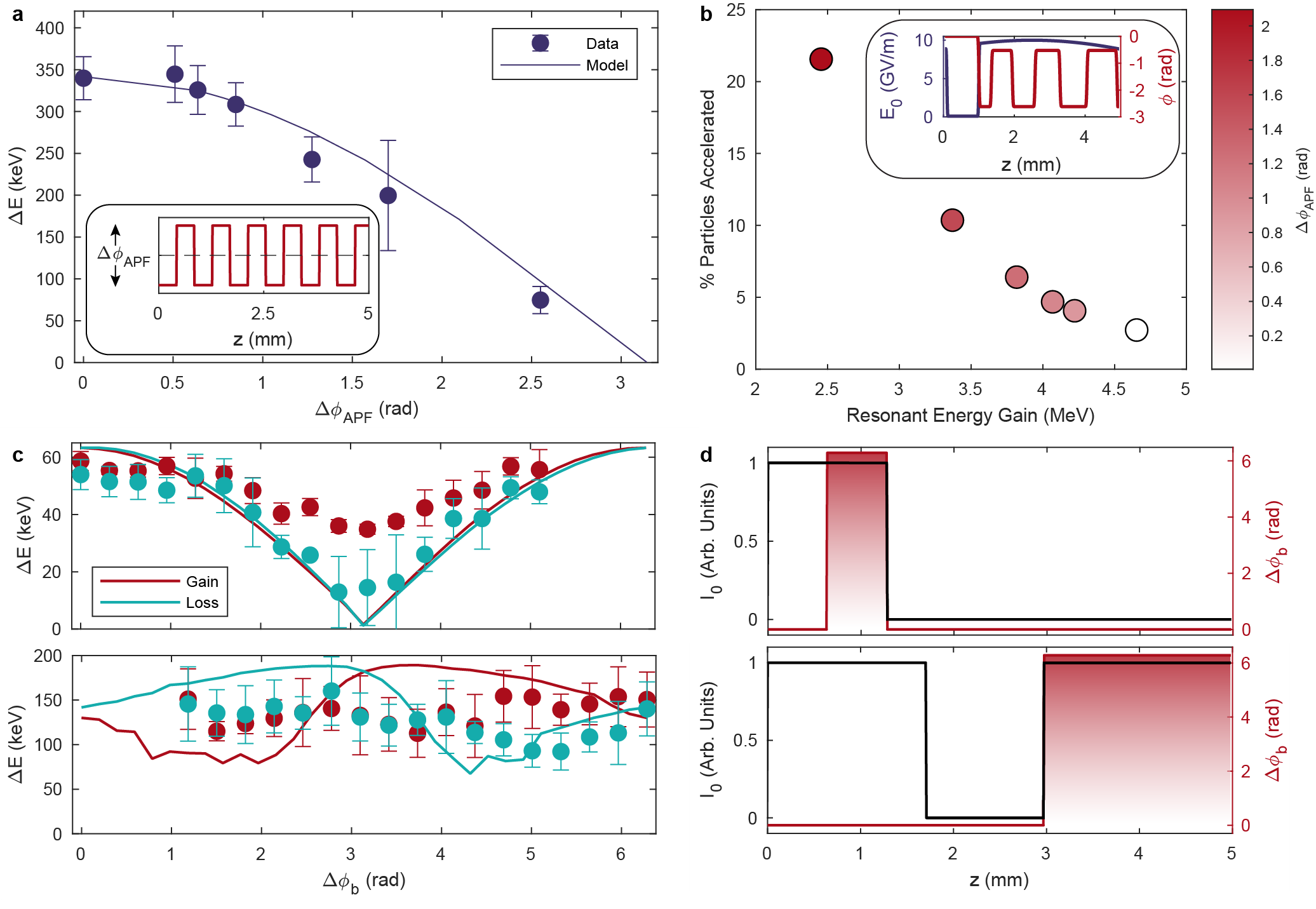} 
    \caption{\textbf{a} Energy gain as a function of $\Delta\phi_{apf}$, shown in the inset. \textbf{b} Simulation results showing the proportion of electrons which reach $0.9\gamma_{res}$ as a function of $\Delta\phi_{APF}$, which directly corresponds to $\Delta{E}_{res}$. The inset shows an example phase and amplitude input with a 1~mm buncher followed by APF confinement and acceleration.\textbf{c} Top panel shows the energy gain vs. $\Delta\phi_b$ for a 1.28~mm interaction, where the first 0.64~mm is accelerated at the matched phase, and a phase jump $\Delta\phi_b$ is added for the second 0.64~mm. In the bottom panel, an initial 1.7~mm section is used to accelerate the beam followed by  a 1.28~mm drift section and the rest of the 5~mm structure used to shift the energy spectrum for different $\Delta\phi_b$. At top (bottom), the beam has starting energy of $\gamma = 17 (12)$. Error bars show the standard deviation of 5 shots at each $\phi_b$. Solid lines indicate simulation results. \textbf{d} Phase and amplitude for scans shown in \textbf{c}. }
    \label{fig:Dynamics}
\end{figure}

In an APF scheme, particles are continuously accelerated at a constant, often large, fraction of the maximum possible gradient proportional to $\cos(\phi_{APF})$ where $\phi_{APF}$ is the synchronous phase. The phase actually jumps repeatedly between $\pm \phi_{APF}$ so that field experienced by the particles alternates between longitudinally focusing, transversely defocusing and longitudinally defocusing, transversely focusing regions. Overall, this forms a so-called FODO lattice which can stably confine a particle beam. In our setup, we test the APF concept by adding to the matched SLM phase profile a series of discrete phase jumps of amplitude $\pm \phi_{APF}$ where the sign switches twice each cell length $\lambda_{APF}$  as pictured in the inset on Fig. \ref{fig:Dynamics}a. We plot the maximum observed acceleration vs. the step height $2\phi_{APF}$. In this dataset, $\lambda_{APF} = 0.85$~mm, so every 0.425~mm there is a phase jump. The observed energy modulation aligns well with our model, showing again that we are able to coherently accelerate a particle bunch at a programmed resonant phase. By increasing $\Delta\phi_{APF}$ up to $\pi$, we almost completely eliminate energy gain in this multi-mm interaction. 

In order to increase not just the number of particles transmitted, but also the quality of the accelerated beam, it is important to prepare the longitudinal phase space of the beam to avoid injection over all phases of the accelerating optical wave \cite{sears2008production, sun2023high}. This can be done by controlling the amplitude of the fields in the DLA channel and introducing a drift region where an initial energy modulation can quickly convert into density modulation, bunching the beam. To demonstrate the principle, we first illuminate a 1.28~mm region of the structure, with the first 0.64~mm illuminated at constant phase (Fig. \ref{fig:Dynamics}d). When we vary the phase of the latter half, $\Delta\phi_b$, we see that at a $\pi$ phase shift, much of the energy gain and loss on the 8.7~MeV electron beam is canceled out, as we would expect in a ballistic case where phase evolution is minimal (Fig. \ref{fig:Dynamics}c-d, top). The imperfect cancellation can be attributed to varying phase slippage for different particles. In this case, there is no clear shift in the energy spectrum. By lowering the injection beam energy to 6.2~MeV and introducing a drift section, the results are very different. Due to increased dynamic dephasing, at no point in a scan over $\Delta\phi_b$ does the energy gain come close to fully canceling. Instead, the gain and loss regions of the spectrum are shifted depending on $\Delta\phi_b$ (Fig. \ref{fig:Dynamics}c-d, bottom). Comparison with simulations indicate that this shift is due to an induced sub-wavelength density modulation which can be exploited for significantly higher quality output electron beams with decreased energy spread.

\section{Discussion}   

In summary, we have shown that by utilizing a combination of a PFT illumination scheme with an SLM, it is possible to fine tune the phase resonant condition in the DLA accelerator and significantly increase the interaction length and energy gain with respect to previous DLA implementations. We demonstrated acceleration for distances larger than 4~mm, with as much as 0.55~MeV energy gain. The flexible setup also allowed us to test out different focusing schemes; further investigation with better matched injection beams should be performed to test how much particle confinement increased utilizing these schemes. All of this is possible due to the immense tunability of the accelerating laser field, enabled not by complicated custom nanofabricated structures but instead by commercial SLM technology.

The potential for future experiments is showcased in simulations that combine the phase and amplitude control schemes demonstrated in our experiment in an ideal setup where we assume a matched input beam, a 10 GV/m incident field and a significantly larger structure factor up to 0.12 (see supplementary). Fig. \ref{fig:Dynamics}b shows the results of these simulations where the beam is focused and bunched using an APF scheme indicating that over 20\% of the particles entering the structure could be accelerated up to 2.4~MeV in only 5~mm, 1~mm of which is used entirely for bunching. We can see the large effect that APF focusing is having on particle throughput in these simulations as lowering $\Delta\phi_{APF}$ greatly increases the energy gain at the expenses of particle throughput. 
In the Supplementary Material section we also present an ideal simulation case for the ponderomotive focusing scheme. Comparing the output of these simulations, it is clear the potential for APF to provide both higher acceleration gradients and more particle throughput for similar incident laser intensities. Both approaches were experimentally investigated in our work as we note that the ponderomotive focusing scheme is implemented via a smoother phase profile with fewer phase jumps, making it more compatible with LCM-based optical control methods.

The setup presented here enabled rapid testing of the most essential elements of creating a scalable accelerator. The soft, live control enabled by the SLM allows for the tuning of beam energy, transmitted beam current, and energy gain as required by the application \cite{england_schachter, adiv2021quantum, koyama2014parameter}. The scheme is also an almost perfect canvas for utilizing machine learning optimization techniques \cite{edelen2020machine}. In fact, the methodology used here allows for the manipulation of the electron phase space without the need for precise knowledge of either the structure or laser parameters as it allows to tune for the matched phase irrespective of the actual cause of dephasing. When scaling to longer structures and larger energy gains, fabrication errors in structure or phase inhomogeneity in the laser will inevitably stack up and this independence of the tuning from the cause of error will only become more crucial. Programmable optical components allow for arbitrarily complicated phase and amplitude structures to be tested, which could also help to bridge the gap between the sub-relativistic and relativistic regimes, which is critical for the development and future operation of on chip accelerators. 

\section{Methods}
\label{sec:methods}
\begin{figure}
    \centering
    \includegraphics[width=\textwidth]{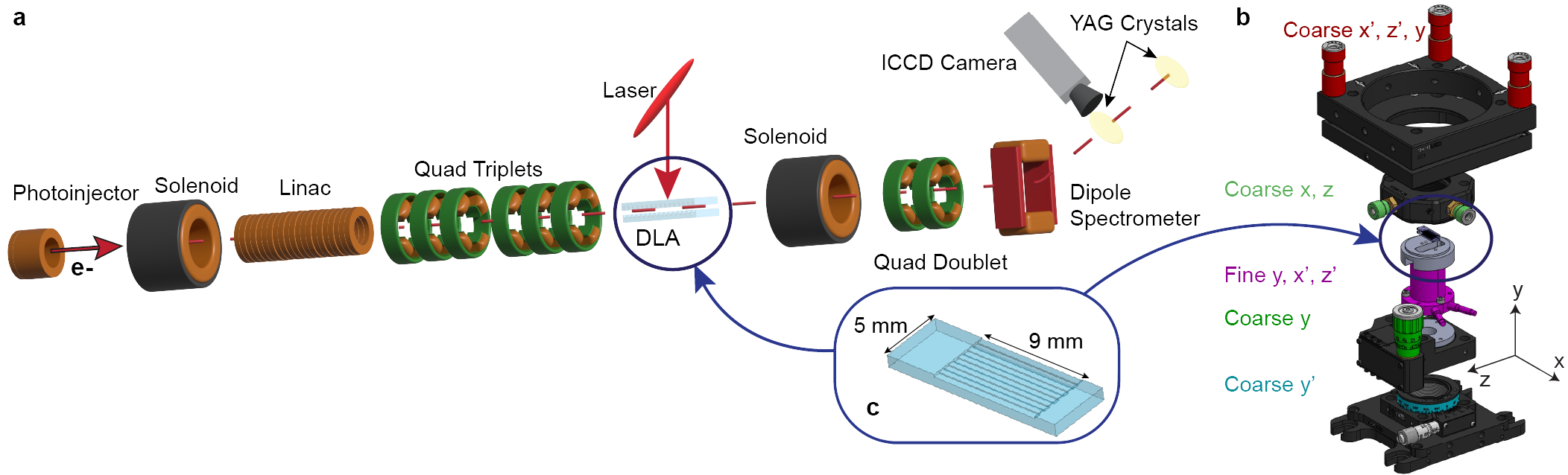}
    \caption{Schematics \textbf{a} Pegasus beamline at UCLA. The electrons are photoemitted in the gun, then accelerated in the linac and then focused by quadrupole magnets into the DLA structure. After the laser interaction they are transported to the dipole spectrometer where the energy spectrum can be observed on one of two YAG crystals using an ICCD camera. \textbf{b} Exploded view of the structure assembly. \textbf{c} View and dimensions of one of the two gratings in the structure.}
    \label{fig:supplementarySchematics}
\end{figure}

\subsection{Electron Beam Measurement}
\label{sec:beamline}
For this experiment, the electron beam has a charge ranging from 0.2 to 5~pC and is first accelerated by a radiofrequency photoelectron gun before being boosted by the LINAC to a total energy of 8.7 MeV (6.2 MeV for Fig. \ref{fig:Dynamics}b) (Fig. \ref{fig:supplementarySchematics}). It is then focused by a series of 6 quadrupoles to a spot size of $\sigma_x = 23 \, \mu\text{m} \times \sigma_y = 54 \, \mu\text{m}$ at the the DLA plane. The transmitted beam is then transported by a solenoid and additional quadrupoles to the end of the beamline, where the energy spectrum is viewed by an ICCD camera after a dipole spectrometer on one of two screens. The closer screen has a dispersion of 10~cm while the far screen has a dispersion of 32.8~cm; the far screen thus has higher energy resolution but a smaller spectral field of view. The close screen has a spatial resolution of 44~$\mu$m/pixel, and we therefore find a calibration of 3.85~keV/pixel for the close screen while operating at a total energy of 8.7~MeV, as in the case of the maximum gain reported in Fig. \ref{fig:phase_basic}

Similarly to ref. \cite{crisp2024extended}, the maximum energy gain and loss are extracted from the spectrometer profiles by defining a figure of merit defined as $FOM$ = $\frac{\langle S_1\rangle- \langle S_0\rangle}{\sigma_0}$, where $\langle S_{1 (0)} \rangle$ are the observed spectra averaged over at least 5 laser on (off) shots and $\sigma_0$ is the standard deviation of the laser off shots. To determine the maximum gain/loss at the tails of the spectrum where the modulated electron density nearly blends into the noise floor, we require at least 3 consecutive pixels to have a signal to noise ratio of at least $FOM > 1.5$.

For the ponderomotive focusing datasets, the high dispersion spectrometer screen was used to increase energy resolution. As a consequence of the relatively low energy acceptance, there is cutoff on the maximum energy modulation observable for the central $n = 0$ peak. However, by increasing the structure gap and decreasing the incident field strength, we are still able to verify that the relative excitation amplitude of the $\pm$1 modes agree well with our model as shown in Fig. \ref{fig:Ponderomotive4}b. 

\subsection{Structure}
\label{sec:methods_structure}
The structure is composed of two fused silica gratings of constant 780~nm periodicity. The gratings were designed and simulated using the finite-difference time-domain software Lumerical to provide the optimal field structure for long interactions within the limits of single side illumination \cite{crisp_asymmetricDesign}. While a dual drive has shown improved control on the symmetry of the fields in the gap, for experimental simplicity we elect to operate with a single laser drive and tuned separately the teeth height of each grating to compensate for the asymmetric power delivery and minimize the transverse kick of the DLA structure. 

The upper grating has a tooth height of 450~nm, reducing the diffraction efficiency compared to the lower, 650~nm grating. This symmetrizes the fields within the structure for certain gap-offset combinations, resulting in an accelerating cosh mode. The gratings are 1~mm thick, 5~mm long and 9~mm wide. The thickness of the grating serves as a filter; any electrons transmitted to the spectrometer screens pass through the DLA gap. The gratings are independently mounted with a gap between them adjustable over a range of 10~$\mu$m using a 3 piezo motor stage (Fig. \ref{fig:supplementarySchematics}).

The structure is assembled using diode laser interferometry to ensure sub-micron flatness as described in \cite{Crisp2022_structureChar}. In the same way, the relative rotation is tuned with sub-microradian accuracy. The full assembly is mounted in vacuum on a stage with motors controlling the $x$, $y$, $x'$, and $y'$ axes. In-situ transmission scans which measure particle count as a function of piezo value confirm that unless otherwise noted, the structure gap was equal to 1.2~$\mu$m. For our input electron beams, this results in only 1.6\% transmission, so that the charge collected at the spectrometer screen is typically less than 1~fC. When the gap size is set to 1.2~$\mu$m, using the energy modulation induced by a short, non-PFT laser pulse, we measure the structure factor to be $\kappa = 0.05$ in agreement with the simulation predictions. 

The implementation of dual illumination or a Distributed Bragg Reflector would greatly improve the acceleration gradient. As discussed in Crisp et al. \cite{crisp_asymmetricDesign}, asymmetric illumination of a symmetric structure introduces a deflection force, which for high gradient interactions results in particle loss. As a result, in this experiment we deliberately lowered the design structure factor to enable long interactions, and utilized a larger than wavelength gap for the same purpose. Dual illumination would allow for the use of higher efficiency gratings and a smaller gap size, both of which increase structure factor; for example the structure used by Peralta et al. \cite{Peralta2013} had a structure factor of up to 0.24, but there was no stable accelerating channel. 

\subsection{Drive Laser}
\label{sec:methods_laser}
A 20~mJ, 780~nm, 100~fs FWHM Ti:sapphire laser is split 9:1 with the low energy arm redirected to a frequency tripling path used to generate the UV which drives the photocathode in the RF gun. The majority is used to drive the DLA, with a 2:1 telescope immediately before the 600~ln/mm PFT generating grating \cite{fulop2010applications}. The laser is incident on the grating at a $27.9^{\circ}$ angle so that the diffracted beam is along the grating normal in order to minimize temporal distortions over the beam profile \cite{Kreier2012_OpticalDistortions}. A 2-lens telescope images the grating onto the SLM with a magnification of $M = 0.55$; another 2 lens system with $M = 0.85$ images the SLM onto the DLA. These specific magnifications were chosen to best decouple the PFT angle $\theta_{PFT}$  and pulse length $\tau$. The total magnification is defined by the electron velocity, $v = \beta{c}$, such that $\beta = \cos(\theta_{PFT})/\sin(\theta_{PFT} + \theta_I)$ is satisfied. Since $\lambda_0 = \lambda_g$, $\theta_I<5$ mrad, and $\theta_{PFT} = 45^{\circ}$ is required for group velocity matching. 

The optical system is designed with 3 imaging planes: at the location of the 600~ln/mm diffraction grating used to generate PFT, at the SLM plane, and at the final DLA plane. The PFT angle is verified using an interferometric measurement between the PFT laser and a copropagating reference pulse \cite{Cesar2018_opticalDesign}. The use of 2 lenses between the PFT grating and SLM allows the PFT angle to be (mostly) decoupled from the imaging condition, allowing for the pulse to remain short whilst the PFT angle is tuned. A cylindrical lens in the second stage means that we do not image in $x$, allowing the Fresnel lens phase mask to modulate the intensity of the laser while technically operating the SLM in a phase only configuration.

\subsection{Spatial Light Modulator}
\label{sec:methods_SLM}
We use a commercial liquid crystal on silicon reflective SLM with a 1920~x~1080 pixel array with $>93\%$ fill factor. The optical axis is at 45$^\circ$. We use a half wave plate before and after the SLM such that the laser polarization is on the same axis so we are in a phase-only modulation configuration. The laser is incident at a 10$^\circ$ angle from the normal to the liquid crystal panel, introducing minimal phase errors. The SLM has a 12.5~mm by 7.1~mm panel active area which the laser overfills transversely, corresponding to the $z$ electron propagation axis, but not vertically. The SLM has a maximal 75\% reflectance, and a theoretical diffraction efficiency of above 50\% for the phase masks used. The total energy throughput of the optical transport is limited by the 138~mJ/cm$^2$ damage threshold of the SLM.

The SLM does not have enough phase range to encompass the entire desired mask, causing jumps along the laser profile at all points where the phase is reset. Unfortunately, imaging the first order reflection of the SLM onto the DLA plane implies that at each phase jump there is a corresponding gap in intensity (Fig. \ref{fig:Ponderomotive4}\textbf{a}) leading to a degradation of the DLA performance. Many of the phase jumps are due to the optimization of the resonant phase profile, but some more are introduced by the implementation of the ponderomotive and APF focusing schemes. Interestingly, due to the gentler phase variation, the ponderomotive focusing scheme is only responsible for a minimal number of these phase jumps, while for APF this is a more serious issue which is exacerbated by the requirement to propagate the laser through 1~mm of fused silica. Non linear effects at the points where there are large variation in intensity result in large nonlinear phase shifts posing significant challenges to retain phase coherence for the full structure length. Thinner substrates could greatly mitigate this problem.

\subsection{Spatial and Temporal Alignment}
\label{sec:methods_electrons}
Spatial alignment is obtained by viewing a 45$^\circ$ phosphor screen with an overhead CCD camera. The electron beam is focused to a small spot and then transmitted through the structure. The tunability of the structure allows us to maximize transmission by tuning the pitch with a 10~$\mu$m gap and then closing the gap to the optimal run condition of 1.2~$\mu$m. When lower structure factors are desired the structure gap can be increased.

Preliminary timing overlap is obtained by placing a photodiode at the interaction plane. A delay stage prior to the main optical table to scan the relative time-of-arrival in 16~fs steps, much lower than the decay time of the photodiode. Femtosecond scale synchronization is obtained using a reference laser pulse without any pulse front tilt such that the bulk of the electron beam is energy modulated by the DLA, even if only for the 100~fs spatial extent of the laser pulse. After this, an unmatched grating (900~ln/mm) is used to generate $\theta_{PFT} = 56^{\circ}$ to establish electron energy gain and verify temporal overlap, again with a limited interaction length, before swapping to the final 600~ln/mm grating which provides $\theta_{PFT} = 45^{\circ}$.

\subsection{Simulation}
\label{sec:Simulation}
Simulations are done using a modified version of the Matlab-based numerical code SHarD by Ody et al. \cite{Ody2021}. The code accepts as input arbitrary phase and amplitude profiles for the laser and decomposes the field into a large number (typically $>$ 100 different harmonics, then it uses a Runge-Kutta method to update the particle position and momentum. The particles can be traced throughout their transport, and are eliminated when they hit the structure walls. The structure can be defined simply by a gap, length, and structure factor, or by implementing a full complete model of the diffraction from the top and bottom gratings using complex coefficients \cite{Black2020}.

To model our experimental setup, we use a 5000 particle electron beam with a normalized emittance of 200~nm, $\sigma_y = 30~\mu$m, and total energy $6.1 (8.7)$~MeV and 0.05\% energy spread. To save on computing time, we track only the 1.6\% of particles initially within the 1.2~$\mu$ aperture of the DLA. For the simulations in Fig. \ref{fig:Concept}c-d and Fig. \ref{fig:Dynamics} we use a more idealized electron beam with 1~nm normalized emittance, $\sigma_y = 1~\mu$m, and total energy of $6.1$~MeV. For all simulations, particles are assumed to be injected at uniformly random phase over longitudinal phase space. The bunching section in the idealized simulation is designed to maximize the number of particles trapped at the accelerating phase at $z = 1$~mm. The ponderomotive design follows the stability criterion laid out by Naranjo et al. \cite{Naranjo2012_Ponderomotive}, and the APF lattice design follows the method from Niedermeyer et al. \cite{Niedermayer2018_APFconcept}. In both cases, optimal focusing requires the periodicity of the added phase mask to increase as the energy of the beam increases.

\backmatter


\bmhead{Acknowledgements}
Funding from Gordon and Betty Moore Foundation and DOE grant No. DE-SC0009914



\end{document}